\documentclass[twocolumn,showpacs,preprintnumbers,amsmath,amssymb]{revtex4}
\usepackage{graphicx}% Include figure files
\usepackage{dcolumn}% Align table columns on decimal point
\usepackage{bm}% bold math
\begin{document}

\preprint{R. Palai {\it et al}.}

\title{The $\beta$ Phase of Multiferroic Bismuth Ferrite and Its
$\gamma$-$\beta$ Metal-Insulator Transition}% Force line breaks with \\

\author{R. Palai$^{1}$, R.S. Katiyar$^{1}$, H. Schmid$^{2}$, P. Tissot$^{2}$,S. J. Clark$^{3}$,
J. Robertson$^{4}$, S.A.T. Redfern$^{5}$ and J. F. Scott$^{5}$}
\affiliation{$^{1}$Department of Physics, University of Puerto
Rico, San Juan, PR 00931-3343, USA}

\affiliation{$^{2}$ Department of Inorganic, Analytical and
Applied Chemistry, University of Geneva, CH-1211 Geneva 4,
Switzerland} \affiliation{$^{3}$ Department of Physics, Durham
University, Durham DH1 3LE, UK} \affiliation{$^{4}$ Department of
Engineering, University of Cambridge, Cambridge CB2 1PZ, UK}
\affiliation{$^{5}$ Department of Earth Science, University of
Cambridge, Cambridge CB2 3EQ, UK}
%\email{rkatiyar@uprrp.edu}

\date{\today}% It is always \today, today,

\renewcommand{\baselinestretch}{} % ... as the instructions say
\newcommand{\bfo}{BiFeO$_{3}$}
\newcommand{\tn}{$T_{\rm N}$}
\newcommand{\pr}{$P_{\rm r}$}
\newcommand{\mr}{$M_{\rm r}$}
\newcommand{\tc}{$T_{\rm c}$}
\newcommand{\jc}{$J_{\rm c}$}
\newcommand{\rs}{$R^{\rm 2}$}% for R square
\newcommand{\dc}{$^{\circ}$C}% for degree centigrade
\newcommand{\rr}{$r_{\rm R}$}% for resistance ratio

\begin{abstract}
{\sf We have carried out extensive experimental studies, including
differential thermal analysis, polarized and high-temperature
Raman spectroscopy, high-temperature X-ray diffraction, optical
absorption and domain imaging, and show that epitaxial (001) thin
films of multiferroic bismuth ferrite (\bfo) are monoclinic at
room temperature instead of tetragonal  or rhombohedral like bulk
as reported earlier. We report a orthorhombic order-disorder
$\beta$-phase between 820 and 950 ($\pm$ 5)\dc\ contrary to the
earlier report. The transition sequence rhombohedral-orthorhombic
transition in bulk (monoclinic-orthorhombic in (001)\bfo\ thin
film) resembles that of BaTiO$_{3}$ or
PbSc$_{1/2}$Ta$_{1/2}$O$_{3}$. The transition to the cubic
$\gamma$-phase causes an abrupt collapse of the bandgap toward
zero (insulator-metal transition) at the orthorhombic-cubic
$\beta$-$\gamma$ transition around 950\dc. Our band structure
models confirm this metal-insulator transition, which is similar
to the metal-insulator transition in
Ba$_{0.6}$K$_{0.4}$BiO$_{3}$.}
\end{abstract}

\pacs{64. 70. Kb, 71. 30. +h, 78. 30. - j, 77. 55. + f, 77. 80. Bh }% PACS, the Physics and Astronomy
\maketitle
%\section{Introduction}

Magnetoelectric (ME) multiferroics are technologically and
scientifically important because of their potential applications
in data storage, spin valves, spintronics, quantum electromagnets,
microelectronic devices {\it etc}.
\cite{fiebig:nature1,wang:science1, tokura:science06} and the
novel mechanism that gives rise to electromagnetic coupling.
Ferroelectricity originates from off-center structural distortions
(d$^{0}$ electrons) and magnetism is involved with local spins
(d$^{n}$ electrons), which limits the presence of off-center
structural distortion \cite{spaldin:jpcb}. These two are quite
complementary phenomena, but coexist in certain unusual
multiferroic materials. \bfo\ (BFO) is one of the most widely
studied multiferroic material because of its interesting ME
properties {\it i.e} ferroelectricity with high Curie temperature
(\tc~$\approx$~810-830~\dc) \cite{Fischer:jpc80} and
antiferromagnetic properties below \tn $\approx$~370~\dc\
\cite{Fischer:jpc80,smolenkii:SPU82}. The bulk BFO single crystal
shows rhombohedral ($a$~=~5.58~\AA\ and $\alpha$~=~89.5$^{0}$)
crystal structure at room temperature (RT) with the space group
R3c and $G$-type antiferromagnetism
\cite{smolenkii:SPU82,hans:ac90}. If BFO were an ordinary
antiferromagnet, the space group would be R3m \cite{hans:ac90}.
The structure and properties of bulk BFO have been studied
extensively \cite{Fischer:jpc80, smolenkii:SPU82, hans:ac90,
bucci:jpc72} and although early values of polarization were low
due to sample quality, \pr~=~40~$\mu$C/cm$^{2}$ is now found in
bulk by several different groups \cite{shuartsman:apl07}. The weak
ferromagnetism at room temperature occurs due to the residual
moment from the canted spin structure \cite{smolenkii:SPU82}. It
is very difficult to grow a high quality (defect free and
stoichiometric) bulk single crystal with low leakage, which is
detrimental to the practical applications of this material.

It has been found that thin films of BFO grown on
(100)~SrTiO$_{3}$ (STO) substrates show very high values of \pr\
($\sim$55, 86, and 98 $\mu$C/cm$^{2}$ for the (001), (101), and
(111) BFO films, respectively) and magnetization (\mr$\sim$150
emu/cc) \cite{wang:science1,li:apl04,das:apl06}. (Very recently,
Ricinschi {\it et al}. \cite{ricinschi:jpcm06} have claimed \pr\
of 150~$\mu$C/cm$^{2}$ for the polycrystalline BFO films grown on
Si substrates, but this is probably an artifact due to leakage and
charge injection.) This makes BFO as one of the potential
materials for the novel device applications, although the
mechanism(s) behind the huge polarization claimed by some groups
is not yet fully understood. Some experimental results
\cite{wang:science1,li:apl04} and theoretical reports
\cite{ederer:prb05} suggest that the epitaxial strain might be the
cause of such high value of \pr\  and \mr. However, a recent study
showed that the epitaxial strain does not enhance \mr\ in BFO thin
films \cite{Eerenstein:science05}. It is believed that the
heteroepitaxy induces significant and important structural changes
in BFO thin films, which may lead to very high values for \pr.

There are some controversies in literature about the crystal
structure of (001) epitaxial thin films. There have been several
reports claiming tetragonal \cite{wang:science1,singh:prb05},
rhombohedral \cite{das:apl06, qi:apl05}, and monoclinic
\cite{xu:apl05} structure of (001) BFO films on STO substrates.
Therefore, the sequence of transitions is poorly understood, and
there is no understanding of the overall physics of the phase
transitions involved; more structural analysis of thin films is
necessary for better understanding of engineered epitaxial and
hetrostructure BFO thin films. In the present work we used a
variety of techniques which combined show a sequence of
transitions resembling the well-known 8-site model of barium
titanate.

ABO$_{3}$ oxide perovskites which are rhombohedral at low
temperatures, such as LaAlO$_{3}$, PrAlO$_{3}$, or NdAlO$_{3}$
\cite{scott:pr69, geller:prb70} have ferroelastic instabilities at
the A-ion site that induce displacive phase transitions directly
to cubic; but those which have B-site instabilities instead have
order-disorder transitions to cubic that involve two or more
steps.  This has been successfully described \cite{comes:ssc68,
chaves:prb76} by an eight-site model in which the B-ion
displacements are always locally toward a [111] axis, but
thermally average via hopping over [111], [$\bar{1}$11],
[1$\bar{1}$1], and [11$\bar{1}$] to give orthorhombic, tetragonal,
or cubic time- and space-global averages.  In the present work we
show that this model describes BFO, contrary to conventional
wisdom \cite{haumont:prb06} but in agreement with NMR, which shows
some B-site disorder \cite{blinc:private}.

\subsection{The $\alpha$-phase}

The rhombohedral (R3c), tetragonal ($P4mm$) \cite{wang:science1},
and monoclinic ($Bb$)\cite{ederer:prb05a} structures of BFO give
rise to 13, 8, and 27 distinct Raman-active modes, respectively,
as listed in the Table~\ref{raman}. For the orthorhombic
distortion in the $\beta$-phase, as discussed below, the
tetragonal entries will remain correct, with only a small
splitting of the {\it E}-modes.
\begin{table}
\begin{center}
\caption{\label{raman}{\sf Selection rules for the Raman active
modes for rhombohedral ($R$), tetragonal ($T$) and, monoclinic
($M$) crystal structures in different polarization configurations
with total number of normal ($N$) Raman modes. The notation
"$<$001$>$up" means unpolarized spectra along the pseudo-cubic
$<$001$>$ direction perpendicular to the substrate. The notation Z
(along $<$001$>$ direction)  and \={Z} are the directions of the
incident and backscattered light, respectively.}}
\begin{tabular}{cccc}
 \cline{1-4}
Scattering  & \  $R$(R3c)   & ~~\ $T$($P4mm$)  & ~~~~~\ $M$($Bb$) \\
geometry& ~~~\ (C$_{3v})$& ~~~~~\ (C$_{4v}$)& ~~~~\ (C$_{s}$)\\
\cline{1-4}

N(Raman)& \  4{\it A}$_{\rm 1}$ + 9$E$& ~~\ 3$A_{\rm 1}$ + $B_{\rm 1}$ + 4$E$ & ~~~~\ 13$A^{'}$ + 14$A^{''}$\\
$<$001$>$up& \  4{\it A}$_{\rm 1}$ + 9$E$& ~~\ 3$A_{\rm 1}$ + $B_{\rm 1}$ & ~~~~\ 13$A^{'}$ \\
Z(XX)\={Z}& \  {\it A}$_{\rm 1}$ and $E$& ~~\ $A_{\rm 1}$ and $B_{\rm 1}$ & ~~~~\ $A^{'}$\\
Z(XY)\={Z}& \  {\it E}& ~~\ No modes& ~~~~\ $A^{'}$\\
Y(ZZ)\={Y}& \  $A_{1}$& ~~\ $A_{1}$& ~~~~\ $A^{'}$\\
Y(ZX)\={Y}& \  {\it E}& ~~\ {\it E} & ~~~~\ $A^{''}$\\
\cline{1-4}
 \end{tabular}
  \end{center}
\end{table}

The Raman data for 300~nm thick (001)BFO thin film  on the
(100)STO substrate in the parallel, Z(XX)\={Z} and perpendicular,
Z(XY)\={Z} polarization configurations reveal strong peaks at 74,
140, 172 and 219 cm$^{-1}$, while weaker peaks were observed at
around 261, 370, 406, 478, 529, 609, 808 (very weak), 946, 1093
cm$^{-1}$. All of these peaks are due to the BFO normal modes of
vibrations and none of them are arising from the substrate,
contrary to the earlier reported Raman measurements
\cite{singh:prb05, singh:apl06}. We verified our results using
target materials and also growing (001)BFO films on different
substrates. The existence of 13 identical peaks in both the
Z(XX)\={Z} and Z(XY)\={Z} polarization configurations confirms the
Raman selection rules for the monoclinic structure (see Table
\ref{raman}) instead of tetragonal or rhombohedral as reported
earlier\cite{das:apl06,qi:apl05, singh:prb05}. This observation
verifies the very recent report of monoclinic structure for the
epitaxial BFO films grown on (001)STO substrates by Xu {\it et
al.} \cite{xu:apl05} studied via synchrotron radiation. As the
spectra along Y(ZZ)\={Y} and Y(ZX)\={Y} were heavily dominated by
the scattering from the STO substrate, the contribution from the
BFO film could not be separated. We looked very closely near the
phase transition Temperatures. Two noticeable changes have been
observed in the Raman spectra: disappearance of all stronger peaks
(74, 140, 171 and 220 cm$^{-1}$) at $\sim$ 820~\dc\ with the
appearance of a few new peaks, and complete disappearance of
spectra around 950~\dc. This temperature behavior implies that the
BFO maintains its room-temperature structure up to $\sim$
820~\dc\, indicating the ferroelectric-paraelectric (FE-PE) phase
transition, in agreement with the earlier investigations on BFO
bulk single crystal and polycrystalline
\cite{Fischer:jpc80,haumont:prb06} samples. No evidence of soft
phonon modes implies that the BFO has an order-disorder,
first-order ferroelectric transition, unlike PbTiO$_{3}$. No
decomposition was observed above 810\dc\ contrary to the earlier
studies \cite{bucci:jpc72}, suggesting that our samples had fewer
defects and dislocations.

\subsection{The $\beta$-phase}

The Raman spectra show that four lines ($\sim$~213, 272, 820 and
918 cm$^{-1}$) persist above 820\dc. In the cubic perovskite phase
no first-order Raman lines are allowed; all ions sit at inversion
centers, and all long wavelength phonons are of odd parity. The
data show that the beta phase from 820\dc\ to 950\dc\ cannot be
cubic as reported earlier \cite{haumont:prb06}. Since our
backscattering geometry with incidence radiation along Z-axis
favors {\it A}$_{1}$ and {\it B}$_{1}$ phonons, four Raman modes
(3{\it A}$_{1}$ + {\it B}$_{1}$) are predicted in the tetragonal
(or orthorhombic) perovskite phase (Table~\ref{raman}), in
agreement with experiment. Small orthorhombic splitting of four
unobserved {\it E} modes is predicted, but these modes are
unobserved in the backscattering geometry. We observe no soft
modes in these studies but some merely line-width increases,
suggesting that the $\alpha$-$\beta$ and $\beta$-$\gamma$
transitions are both order-disorder, compatible with the
eight-site model originally developed by Comes {\it et al}.
\cite{comes:ssc68}, and developed in detail by Chaves {\it et al}
\cite{chaves:prb76}.

\subsection{Thermodynamics of phases}

\begin{figure}[h!]
\begin{center}
\includegraphics [width=0.35\textwidth,clip]{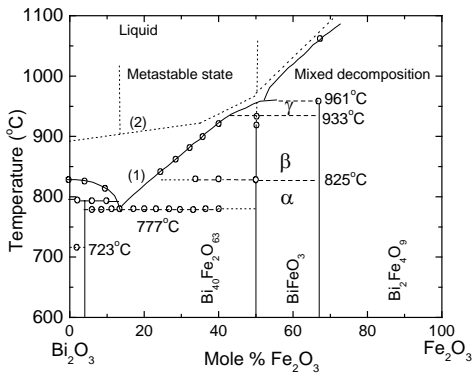}
\caption{\sf Phase diagram of \bfo. Open circles show the data
points obtained by DTA. The DTA peaks were reversible below the
solid line (line (1)) on cooling, while metastable states above
the dotted line (line (2)).  The $\alpha$-phase is monoclinic  in
(001) thin film (rhombohedral in bulk), while the
 $\beta$ and $\gamma$ phases are orthorhombic and cubic,
respectively.} \label{fig:phase}
\end{center}
\end{figure}

\begin{figure}[h!]
\begin{center}
\includegraphics [width=0.35\textwidth,clip]{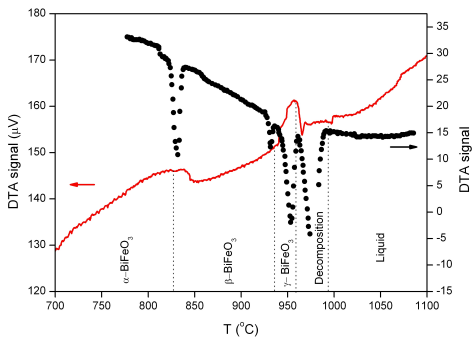}
\caption{\sf DTA studies on crushed single crystal (dotted points)
and thin film target material (solid line) of \bfo. Signal has
been inverted and offset for clarity.} \label{fig:DTA}
\end{center}
\end{figure}

The existence of a $\beta$-phase below the cubic $\gamma$-phase
$\sim$930\dc\ and below the decomposition point at 960\dc\ has
been known for some forty years \cite{Speranskaya:izv65}, but the
early Soviet work is rarely cited, and was sometimes considered
not to be single-phase material. Using differential thermal
analysis (DTA) in conjunction with high-temperature reflected
polarized light microscopy, the BiFeO$_{3}$-Fe$_{2}$O$_{3}$ phase
diagram of Speranskaya {\it et al}. \cite{Speranskaya:izv65} has
been refined (Fig.~\ref{fig:phase}). DTA thermograms shown in
Fig.~\ref{fig:DTA} illustrate the phase transformation sequence
for both \bfo\ single crystals (small crushed dendrites) and thin
film target materials.

These show four sharp endothermic peaks at around 823, 925, 933,
and 961\dc. They can be interpreted as: (a) a first-order
$\alpha$-$\beta$ transition; (b) a $\beta$-$\gamma$ transition;
(c) peritectic decomposition of the cubic phase into flux and
Bi$_{2}$Fe$_{4}$O$_{9}$; and (d) decomposition of
Bi$_{2}$Fe$_{4}$O$_{9}$ into flux and Fe$_{2}$O$_{3}$. The
experimental values obtained on powders of crushed dendrites may
be different for thin films, where transformations may be
controlled by the build up strain energy. These DTA data and
others for which the Bi/Fe ratio is varied produce the phase
diagram shown in Fig.\ref{fig:phase}.

\begin{figure}[h!]
\begin{center}
\includegraphics [width=0.35\textwidth,clip]{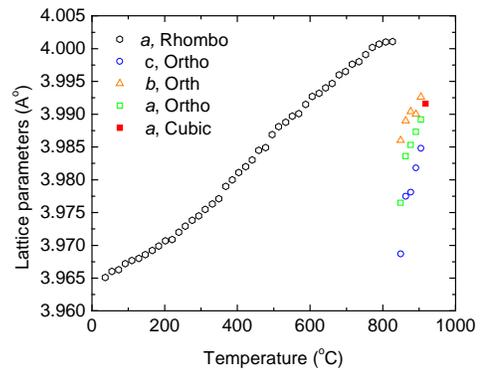}
\caption{\sf Temperature variation of lattice parameters for
rhombohedral (pseudo-cubic setting), orthorhombic and cubic phases
of \bfo. At 825\dc\ the lattice constant {\it a}~=~4.0011~\AA\
splits into a triplet (open circles, squares, and triangles),
which combines again at ca. 925($\pm$ 5)\dc\ in the cubic phase
(solid square), where {\it a}~=~3.9916~\AA.} \label{fig:abc}
\end{center}
\end{figure}
High-temperature X-ray study of BFO powder (Fig.\ref{fig:abc})
showed that the rhombohedral bulk structure has a strongly
first-order transition near 825($\pm$ 5)\dc\ to a P2mm
orthorhombic structure. The unit cell volume shrinks by more than
1\% upon heating through this transition from 64.05 (per formula
unit) in the rhombohedral phase to 62.91 \AA$^{3}$ in cubic. At
around 925($\pm$ 5)\dc\ there is a nearly second-order transition
from orthorhombic to a cubic Pm3m phase. All of the phases have
one formula group per unit cell, as in BaTiO$_{3}$. The
rhombohedral cell parameter is {\it a}~=~4.0011~\AA.  The
orthorhombic cell parameters at 825($\pm$ 5)\dc\ are
$a$~=~3.9765(8)~\AA; $b$~=~3.9860(6)~\AA\ and $c$~=~3.9687(8)~\AA;
and at 905\dc\ $a$~=~3.9892(13)\AA; $b$~=~3.9926(9)~\AA\ and
$c$~=~3.9848(9)~\AA. In the cubic phase $a$~=~ 3.9916(1)\AA\ at
around 925($\pm$ 5)\dc. The shortened bond length in the P2mm and
Pm3m phases, compared with those in the rhombohedral phase, favors
a metallic state. The sequence is similar to the other perovskite
metal-insulator systems such as, NdNiO$_{3}$ and related
rare-earth nickelates showing a sharp decrease of the unit-cell
volume exactly at the metal-insulator transition
\cite{gracia:prb92}.

\subsection{Domains structures in \bfo\ single crystal}

\begin{figure}[h!]
\begin{center}
\includegraphics [width=0.35\textwidth,clip]{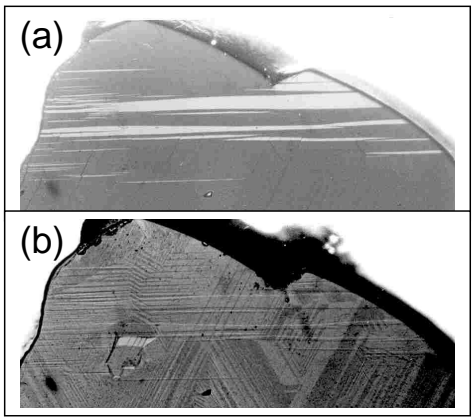}\\
\caption{\sf Evolution of domain structure of $\beta$-phase of BFO
with temperature; (a) at room temperature; (b) at 825\dc\ with
ferroelastic domains.} \label{fig:domain}
\end{center}
\end{figure}

The Raman spectra do not discriminate between orthorhombic and
tetragonal structures for the $\beta$-phase, but domain structures
do. The orthorhombic domains that are symmetry-forbidden in
tetragonal structures are too week to reproduce in
Fig.~\ref{fig:domain}, but in pesudo-cubic (pc) notion. They
unambiguously rule out tetragonal structure for the beta-phase.
The $\beta$-BFO phase has been recognized to be ferroelastic and
non-cubic on the basis of microscopical observation of {\it
ferroelastic} domains using reflected polarized light. The
$\beta$-phase produces rectilinear ferroelastic domains and a
clear phase boundary. On the basis of the ferroelastic domain
pattern of the $\beta$-phase on a (111)pc-cut, showing rectilinear
traces of both (100)pc and (110)pc walls, orthorhombic symmetry
with axes parallel to $<$110$>$ and $<$001$>$ (analogous to
BaTiO$_{3}$) is deduced, whereas a tetragonal phase would allow
only (110)pc walls.
The postulation of a cubic $\gamma$-phase in a very narrow
temperature interval is based upon the reflected polarized light
optical observation that the ferroelastic domains of the
$\beta$-phase disappear at about 925\dc, leaving the sample
optically isotropic up to the decomposition point.  Both types of
domains can be seen in the photographs taken, but one cannot see
rectilinear domain walls in the $\beta$-phase on this photo
(Fig.~\ref{fig:domain}). The ferroelastic nature of the
$\alpha$-$\beta$ phase transition is necessary and sufficient to
satisfy the Toledano \cite{toledano:at74} requirement that changes
in crystal class are required for ferroelastic transitions (this
treats rhombohedral and trigonal as a single super-class). Thus,
for example, triglycine sulphate (TGS) with its
monoclinic-monoclinic C$_{2h}$-C$_{2}$ transition cannot be
ferroelastic.

\subsection{The $\gamma$-phase}

Visual observation shows \cite{hans:fe84} that bismuth ferrite
single crystals at room temperature are yellow and transparent.
The bandgap of BFO is calculated to be $E$$_{g}$~=~2.8~eV
\cite{clark:apl07} and usually measured experimentally as 2.5 eV.
However, it has a large bandgap shift at the $\alpha$-$\beta$
phase transition temperature and turns deep red. If we define the
band edge as where the absorption is 100 in 10~$\mu$m, then the
experimental gap is 2.25, 2.18, 2.00 and 1.69~eV at 20, 160, and
300\dc, respectively. But it shifts abruptly to 1.69~eV at 500\dc,
above which the {\it E}$_{g}$ values are not yet known. There has
been some controversy about its leakage current, with Clark {\it
et al.} \cite{clark:apl07} showing that the ambient bandgap is too
large for intrinsic mechanisms. Optical absorption shows that the
bandgap decreases slowly and linearly with temperature in the
$\alpha$ and $\beta$ phases, from 2.5~eV to ca. 1.5~eV, but then
drops abruptly to near zero at the $\beta$-$\gamma$ transition
near 930\dc. Our observation is thus that $E$$_{\rm g}$ decreases
significantly by the $\alpha$-$\beta$ transition temperature, and
hence conduction in the high-temperature $\beta$-phase can be
intrinsic. In the cubic $\gamma$-phase it is black and opaque.
That is compatible with our band structure calculations (following
section) that a non-cubic distortion is required to give it a
finite bandgap.

Thus we regard $\beta$-BFO phase as semiconducting with small gap
(calculated $\sim$1.5~eV) and $\gamma$-BFO phase from
$\sim$930\dc\ to $\sim$960\dc\ as cubic and metallic; above
960\dc\ decomposition occurs. The insulator-metal transition upon
entering the cubic phase of BFO is very similar to that occurring
in Ba($_{0.6}$)K$_{0.4}$)BiO$_{3}$ \cite{cava:na88, tajima:prb92}
where the material becomes cubic and metallic. This implies in
each case a strong electron-phonon interaction. The similarity
with high-\tc superconducting cuprates in their normal states has
been discussed \cite{sharifi:prl91}.  The data graphed in this
figure were obtained in two different ways:  Below 830~K they were
obtained by conventional absorption spectroscopy at a fixed
temperature; but between 830~K and 1230~K they were obtained at a
fixed wavelength (632.8~nm He-Ne) by slowly varying temperature
and using the Urbach equation to relate absorption coefficient
($a$) to bandgap E$_{g}$(T): $log \ a(T) = (E-E_{g})/E_{g} +
constant$.

\begin{figure}[h!]
\begin{center}
\includegraphics [width=0.35\textwidth,clip]{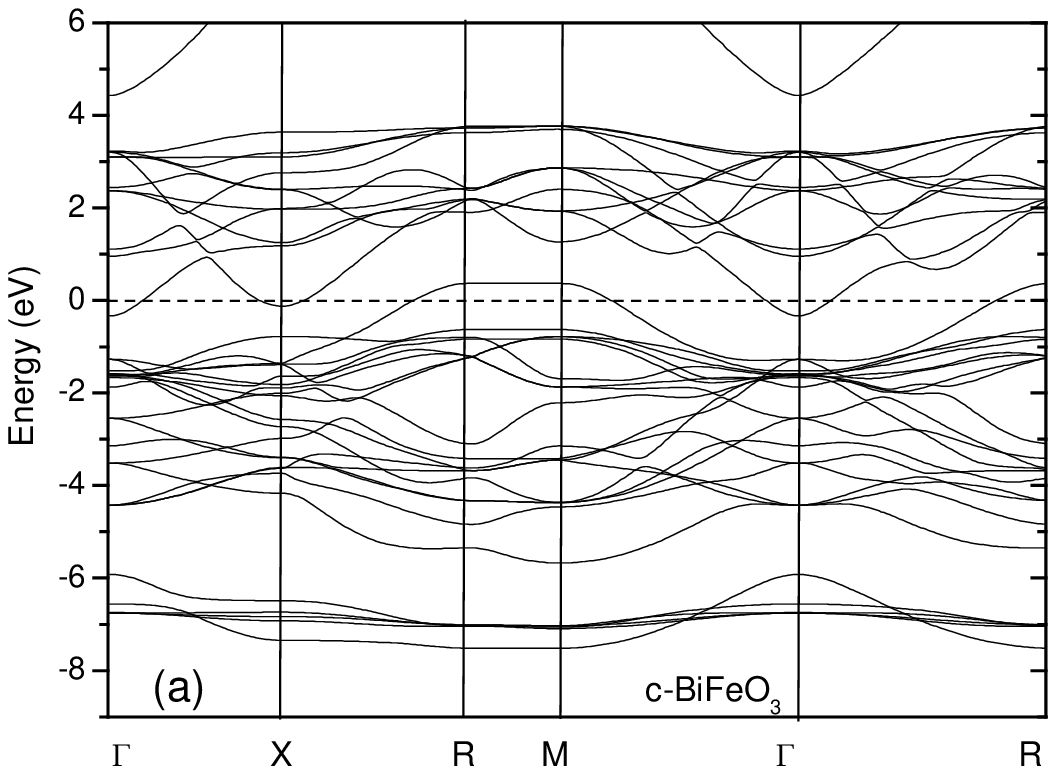}
\includegraphics [width=0.35\textwidth,clip]{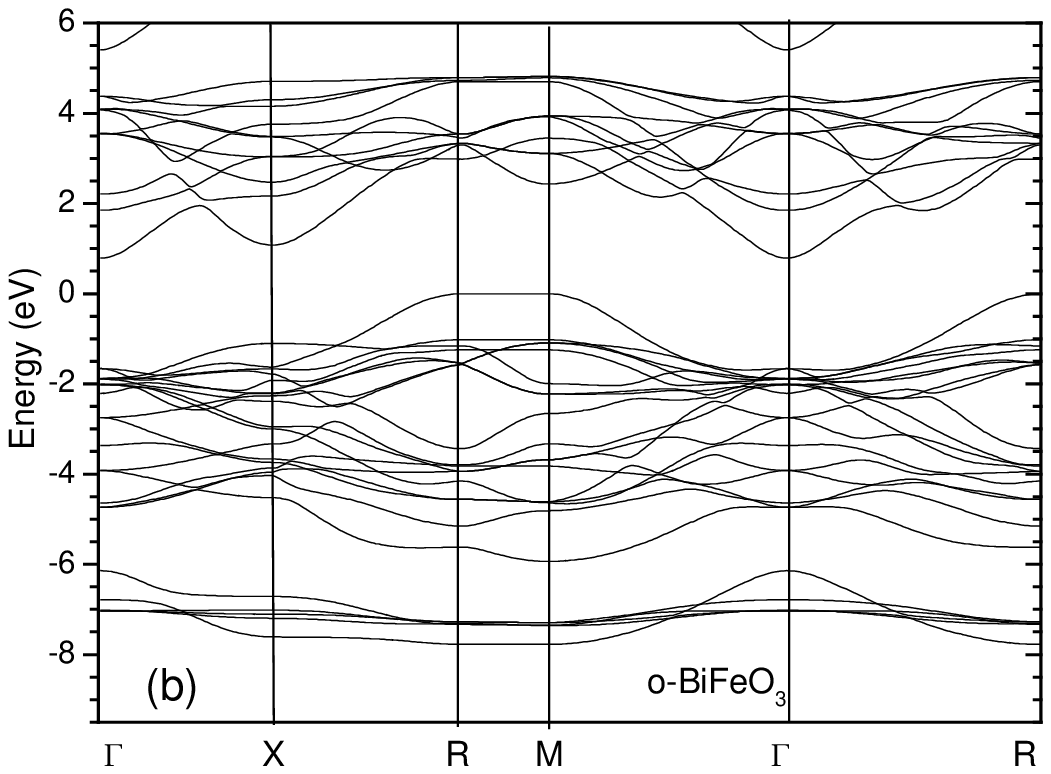}
\caption{\sf Bandgap energy calculation for \bfo\ using screened
exchange method for different structures; (a)cubic; (b)
orthorhombic. Fermi level is at 0~eV in (a) and the valence band
maximum is at 0~eV in (b).} \label{fig:band}
\end{center}
\end{figure}
The differences in transition temperatures between bulk and thin
film, and single crystal, and also between the different studies
drawn together in this paper, show that the transitions in the
thin film are at higher temperatures than in the bulk material.
The effect of pressure on these sorts of transitions is known,
they have a negative dp/dT slope, as pressure stabilizes the
higher symmetry phases in BaTiO$_{3}$. The thin film data would be
consistent with this if you regard any surface relaxation or
epitaxial strains in the thin films as imposing a negative
pressure or lower density on the thin films. The large negative
change in volume observed at the rhombohedral-orthorhombic
transition in BFO is also consistent with the transition having a
strong negative Clapeyron slope, so surface relaxation might be
expected to move the transition to higher temperature in the thin
film.

\subsection{Bandstructure model: Bandgap Collapse and Metal-Insulator Transition}

Obtaining a band gap of zero is a common artifact in theoretical
models using the local density approximation (LDA).  Therefore, in
the present study we circumvent this by using the screened
exchange (sX) method \cite{robertson:pss06}. This method is a
density functional method based on Hartree-Fock, which includes
the electron exchange via a Thomas-Fermi screened exchange term.
It gives the correct band gap of many oxides, including
anti-ferromagnetic NiO and rhombohedral \bfo \cite{clark:apl07}.
We find that the band gap of the cubic phase of BFO is indeed
zero, as shown in Fig.\ref{fig:band}(a).  The Fermi level lies
within the Fe 3d states. It is interesting that in the cubic phase
there is a direct gap at any point in the zone, and this opens up
into a true band gap of 0.8 eV (calculated) in the orthorhombic
phase as seen in Fig.\ref{fig:band}(b). Note that at RT the BFO
has has an indirect band gap, but the direct gap lies only ca.
0.05~eV above it, and the valence band is very flat. Thus, with
increasing $T$, as the conduction band descends, the gap can
become direct.

\subsection{Conclusion}

In conclusion, high quality epitaxial (001)BFO films have been
grown on (100)STO substrates using PLD. The XRD studies showed
that films are $c$-axis oriented with high degree of
crystallinity. The RT polarized Raman scattering of (001)BFO films
showed monoclinic crystal structure contrary to the rhombohedral
and tetragonal as reported earlier. The results obtained from DTA,
high-temperature XRD,  optical absorption, and polarized optics
studies of domains were consistence. We observed the FE-PE
structural phase transition at around 820\dc\; no softening of
Raman modes was observed at low frequencies.  An intermediate
$\beta$-\bfo\ phase between 820-950\dc\ has been observed and
recognized to be {\it orthorhombic} for the first time. The
sequence of monoclinic(film)-orthorhombic-cubic phases or
rhombohedral (single crystal)-orthorhombic-cubic in bulk; the
phase sequence is extremely similar to that in BaTiO$_{3}$.
Moreover, the transitions appear to be order-disorder from the
Raman data, suggesting that the eight-site model of Comes {\it et
al}. \cite{comes:ssc68} and Chaves {\it et al}.
\cite{chaves:prb76} is applicable.  We note that the high-T$_{c}$
superconductor Ba$_{1-x}$K$_{x}$BiO$_{3}$ is also a perovskite
oxide which becomes simultaneously cubic and metallic at x=0.4
\cite{anderson:nac75}. This suggests a similar electron-phonon
coupling; however, in the latter material the Bi-ion is at the
B-site, whereas in \bfo\ it is at the A-site.

\subsection*{Acknowledgement} We thank W. Perez and Dr. M.K. Singh
for experimental help. This work was supported by the DoD
W911NF-06-0030 and W911NF-05-1-0340 grants and by an EU-funded
project "Multiceral" (NMP3-CT-2006-032616) at Cambridge.

\subsection{Method}

BFO thin films of 300~nm thick were grown by pulsed laser
deposition (PLD) using a 248~nm KrF Lambda Physik laser. Films
were grown on STO(100) substrates of area (5~mm)$^{2}$ with $\sim$
25 nm thick SrRuO$_{3}$ (SRO) buffer layer. The growth parameters
were as follows: substrate temperature of 700~\dc, oxygen pressure
of 10~mTorr, laser energy density of 2.0~J~cm$^{-2}$ at a pulse
rate of 10~Hz, and a target-substrate distance of 50~mm. After the
deposition, the chamber was vented with 0.4 atm of oxygen and then
cooled at a rate of 30~\dc/min to room temperature with an
intermediate holding at 500\dc\ for 20 min. The orientation,
crystal structure and phase purity of the films were examined
using Siemens D5000 X-ray diffractometer. The Jobin Yvon T64000
micro-Raman microprobe system with Ar ion laser ($\lambda$ = 514.5
nm) in backscattering geometry was used for polarized and
temperature depended Raman scattering. The laser excitation power
was 2.5 mW and the acquisition time was 10 min per spectrum.  The
single crystal samples were made from crushed dendrites. The
high-temperature X-ray data were collected from powdered bulk
material using a Bruker D8-ADVANCE diffractometer with
Cu-radiation, G$\ddot{o}$bel Mirror, and fast Vantec (linear psd)
detector. Scans from 20 to 80$^{o}$ 2$\theta$ were obtained every
15$^{o}$ from 40$^{o}$ to the decompostion point of BFO with an
acquisition time of 8 min per pattern and a heating rate of
30\dc/min between scans.

\end{document}